\documentclass[%
 reprint,
 amsmath,amssymb,
 aps,
]{revtex4}

\usepackage{graphicx}
\usepackage{dcolumn}
\usepackage{bm}

\begin{document}

\preprint{APS/123-QED}

\title{Indirect Measurement for Optimal Quantum Communication \\ Enhanced by Binary Non-standard Coherent States}

\author{Min Namkung}
\author{Jeong San Kim}%
 \email{freddie1@khu.ac.kr}
\affiliation{%
 Department of Applied Mathematics and Institute of Natural Sciences, Kyung Hee University, Yongin 17104, Republic of Korea
}%

\date{\today}

\begin{abstract}
It is well known that the Helstrom bound can be improved by generalizing the form of a coherent state. Thus, designing a quantum measurement achieving the improved Helstrom bound is important for novel quantum communication. In the present article, we analytically show that the improved Helstrom bound can be achieved by a projective measurement composed of orthogonal non-standard Schr\"{o}dinger cat states. Moreover, we numerically show that the improved Helstrom bound can be nearly achieved by an indirect measurement based on the Jaynes-Cummings model. As the Jaynes-Cummings model describes an interaction between a light and a two-level atom, we emphasize that the indirect measurement considered in this article has potential to be experimentally implemented.
\end{abstract}

\maketitle

\section{Introduction}
Over several decades, optical communication has been widely used for transmitting and receiving important data among distinct parties. The key feature of the optical communication is to discriminate a message encoded in a light with maximal success probability. Unfortunately, the statistical and physical feature of a light is limited, therefore maximal success probability with conventional measurements could not surpass the shot noise limit (equivalently, the standard quantum limit)\cite{j.proakis}. For this reason, large amount of research effort has been contributed to design an unconventional quantum measurement surpassing the shot noise limit in quantum communication\cite{i.a.burenkov0}. 

It is well known that, if one bit message is encoded in one of binary standard coherent states (S-CS), a quantum measurement achieving the Helstrom bound\cite{c.w.helstrom} is described by a projective measurement composed of orthogonal Schr\"{o}dinger cat states\cite{c.w.helstrom2,m.sasaki}. However, the unitary operator for controlling the Schr\"{o}dinger cat states is not Gaussian\cite{m.sasaki}, which makes it difficult to implement the projective measurement via linear optics\cite{g.cariolaro}. Thus, the previous unconventional measurements took the experimental form including displacement operation and detection of photon numbers\cite{r.s.kennedy,m.takeoka} instead of implementing optimal projective measurements. 

In 1973, Dolinar proposed an unconventional measurement with electric feedback, which can achieve the Helstrom bound between binary coherent states\cite{s.j.dolinar}. Based on the Dolinar's idea, many researchers have proposed unconventional measurements for discriminating $N$-ary coherent states\cite{s.j.dolinar2,f.e.becerra,s.izumi,r.s.bondurant} with success probability surpassing the shot noise limit, namely Dolinar-like receivers. Unfortunately, it was also shown that there is a family of $N$-ary coherent states, with which any Dolinar-like receivers cannot achieve the Helstrom bound\cite{k.nakahira}. Thus, it is important and even necessary to consider a new structure of unconventional measurements which beyonds the traditional design including displacement operation, photon number detection, and electric feedback. 

Recently, some indirect measurements for novel quantum communication have been proposed\cite{r.blume-kohout,m.p.dasilva,r.han,r.han2,m.namkung,c.delaney}. Surprisingly, it was shown that an indirect measurement via an interaction between a coherent state and an $N$-level atom is possible to nearly achieve the Helstrom bound\cite{r.han2,m.namkung,c.delaney}, even without electric feedback. Moreover, even if sequential observers perform a quantum communication, the maximal success probability among the sequential observers can nearly achieves the Helstrom bound\cite{m.namkung}.

It should be noted that, not only the methodology for designing an unconventional measurement, the research for finding a suitable optical light also has been performed\cite{i.a.burenkov,i.a.burenkov2,m.t.dimario,r.yuan,s.izumi2,j.dakja,e.m.f.curado}. In this topic, the main purpose was to investigate how much the Helstrom bound can be improved by changing the form of an (ideal or noisy) optical state. For example, the Helstrom bound is known to be improved by squeezing operation on a coherent state\cite{g.cariolaro}. Likewise, a phase-diffused coherent state\cite{m.t.dimario}, a thermal optical state\cite{r.yuan}, and a single rail qubit\cite{s.izumi2} were considered as an information carrier. Recently, it was shown that the generalization of binary coherent states can increase the Helstrom bound\cite{e.m.f.curado}. This result implies that the non-standard coherent state (NS-CS) can enhance the capability of quantum communication. Especially, if binary (modified) Susskind-Glogower coherent states (mSG-CS) are considered \cite{j.-p.gazeau}, Helstrom bound can be equal to one without large mean photon number. Therefore, concerning binary coherent states, finding the optimal structure of the unconventional measurement is important for implenting the enhanced quantum communication. 

In the present article, we propose an optimal measurement for the quantum communication between binary NS-CS in a mathematical manner. We analytically show that the projective measurement can achieve the Helstrom bound and illustrate our result by a projective measurement composed of orthogonal non-standard Schr\"{o}dinger cat states. Although the projective measurement takes the simple form, it is not yet clear how to implement this projective measurement experimentally, except for the case of Glauber-Sudarshan coherent states (GS-CS)\cite{s.izumi3}.\footnote{This topic will be discussed in our future work.} 

To concern the technical or experimental feasibility, we also numerically show that the indirect measurement based on the Jaynes-Cummings model\cite{e.t.jaynes} can nearly achieve the Helstrom bound between binary NS-CS. This implies that the indirect measurement can perform the quantum communication enhanced by the binary NS-CS. Whereas Dolinar-like receivers need real-time electric feedback, our indirect measurement does not. In other words, it is possibly easier to implement our indirect measurement than Dolinar-like receivers. As the experimental implentation of ultrastrong Jaynes-Cummings model is still actively studied\cite{j.-f.huang}, the indirect measurement considered in this article can be expected to be implemented in the near future.

The present article is organized as follows. In Section 2, we briefly introduce the binary quantum communication enhanced by NS-CS. In Section 3, we provide the structure of the optimal meaurement for the binary quantum communication, and numerically show that the optimal quantum communication is nearly performed by an indirect measurement based on the Jaynes-Cummings model. Finally, in Section 4, we propose the conclusion of the present article.

\section{Preliminaries: Enhanced Binary Quantum Communication}
In principle, quantum communication between a sender and an observer can be described as the problem of quantum state discriminaton\cite{g.cariolaro}. The main purpose of the quantum state discrimination is to establish the optimal strategy for discriminating several quantum states with maximal success probability. So far, several quantum state discrimination strategies have been proposed such as minimum error discrimination\cite{j.bae,d.ha,d.ha2}, unambiguous discrimination\cite{i.d.ivanovic,d.dieks,a.peres,g.jaeger} and fixed rate of inconclusive result\cite{j.fiurasek,u.herzog}, as well as suitable measurements for possible implementation\cite{k.banaszek,f.e.becerra2,k.nakahira2}. However, unambiguous discrimination can be realized only in the ideal case\cite{m.namkung3}, and fixed rate of inconclusive result is not theoretically well developed. For these reasons, the experimental implementation of the quantum communication have been mainly focused on the minimum error discrimination.

Throughout this article, we consider the situation that a sender encodes one bit message $x\in\{0,1\}$ into a pure state $|\psi_x\rangle$ in a Hilbert space $\mathcal{H}$ and send it to an observer. In minimum error discrimination, the observer performs a measurement described by a positive-operator-valued-measure (POVM) $\{M_0,M_1\}$, where $M_y$ is a POVM element corresponding to a measurement outcome $y\in\{0,1\}$. The minimum error discrimination is to minimize the following error probability,
\begin{equation}\label{error_probability}
P_e(|\psi_0\rangle,|\psi_1\rangle|q_0,q_1)=q_0\langle\psi_0|M_1|\psi_0\rangle+q_1\langle\psi_1|M_0|\psi_1\rangle,
\end{equation}
where $q_x$ is the prior probability that $|\psi_x\rangle$ is prepared. This minimization problem is equivalent to maximize the following success probability,
\begin{equation}\label{success_probability}
P_s(|\psi_0\rangle,|\psi_1\rangle|q_0,q_1)=q_0\langle\psi_0|M_0|\psi_0\rangle+q_1\langle\psi_1|M_1|\psi_1\rangle.
\end{equation}

The maximal success probability in Eq.~(\ref{success_probability}) is called the Helstrom bound\cite{c.w.helstrom}, which is analytically given by,
\begin{eqnarray}
P_{hel}(|\psi_0\rangle,|\psi_1\rangle|q_0,q_1)&:=&\max_{\{M_0,M_1\}}P_s(|\psi_0\rangle,|\psi_1\rangle|q_0,q_1)\nonumber\\
&=&\frac{1+\sqrt{1-4q_0q_1|\langle\psi_0|\psi_1\rangle|^2}}{2}.\label{helstrom_bound}
\end{eqnarray}
For the case that the prior probabilities $\{q_0,q_1\}$ are fixed, the Helstrom bound in Eq. (\ref{helstrom_bound}) is determined by inner product of $|\psi_0\rangle$ and $|\psi_1\rangle$. Choosing suitable optical states, quantum communication can be enhanced by improving the Helstrom bound. In the present article, we consider generalized coherent states for possible improvement of the Helstrom bound.
\\ \\
\textbf{Definition 1\cite{e.m.f.curado}.} If a pure state takes the form:
\begin{equation}
|\alpha,\vec{h}\rangle:=\sum_{n\in\mathbb{Z}^+\cup\{0\}}\alpha^n h_n(|\alpha|^2)|n\rangle,\label{general_form_coherent_state}
\end{equation}
then the pure state is a \textit{generalized coherent state}. Here, $\mathbb{Z}^+$ is the set of positive integers, $\{|n\rangle|n\in\mathbb{Z}^+\cup\{0\}\}$ is an orthonormal basis of a Hilbert space $\mathcal{H}$, $\alpha$ is a complex number with $|\alpha|<R$ ($R$ can be whether a finite positive number or infinite), and $\vec{h}:=(h_0,h_1,h_2,\cdots)$ is a tuple of real-valued functions $h_n:[0,R^2]\rightarrow \mathbb{R}$ such that
\begin{eqnarray}
&&\sum_{n\in\mathbb{Z}^+\cup\{0\}}u^n\left\{h_n(u)\right\}^2=1,\label{cond1}\\
&&\sum_{n\in\mathbb{Z}^+\cup\{0\}}nu^n\left\{h_n(u)\right\}^2\textrm{ is a strictly increasing function of }u,\label{cond2}\\
&&\int_0^{R^2}duw(u)u^n\left\{h_n(u)\right\}^2=1, \ \ \textrm{for a function }w:[0,R^2]\rightarrow\mathbb{R}^+.\label{cond3}
\end{eqnarray}
Here, Eq. (\ref{cond1}) is for the normalization of Eq. (\ref{general_form_coherent_state}), Eq. (\ref{cond2}) is for the consistent definitions of mean photon number in quantum theory and statistical theory, and Eq. (\ref{cond3}) is for the resolution of an identity operator. 

It is noted that a Glauber-Sudarshan coherent state (GS-CS) is a special case of Definition 1.
\\ \\
\textbf{Definition 2\cite{e.m.f.curado}.} If every real-valued function $h_n(u)$ takes the form:
\begin{equation}
h_n(u)=\frac{1}{\sqrt{n!}}e^{-\frac{u}{2}}, \ \ \forall n\in\mathbb{Z}^+\cup\{0\}\label{h_n}
\end{equation}
then $|\alpha,\vec{h}\rangle$ is so called GS-CS or \textit{standard coherent state} (S-CS). Otherwise,  $|\alpha,\vec{h}\rangle$ is so called \textit{non-standard coherent state} (NS-CS).
\\ \\
\noindent\textit{Remark 1.} The tuple $\vec{h}$ determines both the explicit form of NS-CS and the Mandel parameter $Q_M^{\vec{h}}(u)$, 
\begin{equation}
Q_M^{\vec{h}}:=\frac{(\Delta n)^2}{\langle n\rangle}-1,
\end{equation}
where $\langle n\rangle$ is mean photon number and $\Delta n$ is standard deviation of the photon number. It is well known that the Mandel parameter is directly related with physical properties\cite{e.m.f.curado}: Optical spin coherent state (OS-CS)\cite{a.m.perelomov}, Barut-Girardello CS (BG-CS)\cite{a.o.barut} and modified Susskind-Glogower coherent state (mSG-CS)\cite{j.-p.gazeau,l.susskind} has negative Mandel parameter ($Q_M^{\vec{h}}<0$). This implies that OS-CS, BG-CS and mSG-CS are \textit{sub-Poissoninan}. Meanwhile, Perelomov coherent state (P-CS) has negative Mandel parameter ($Q_M^{\vec{h}}>0$). This implies that P-CS is \textit{super-Poissonian}. As sub-Poissonianity of NS-CS can improve the Helstrom bound\cite{e.m.f.curado}, we mainly focus on the sub-Poissonian NS-CS.

\section{Optimal Measurement for Binary Quantum Communication}

In this section, we propose two kinds of measurements for binary quantum communication with NS-CS. First, we analytically show that an optimal measurement for discriminating NS-CS is described by a projective measurement composed of orthogonal non-standard Schr\"{o}dinger cat states. Moreover, we numerically show that an indirect measurement based on Jaynes-Cummings model is a nearly optimal measurement for discriminating NS-CS.

\subsection{On-off Keying (OOK) and Binary Phase Shift Keying (BPSK) signals}
On-off keying (OOK) and binary phase shift keying (BPSK) signals are well known information carrier in quantum communication\cite{i.a.burenkov0,g.cariolaro}. In OOK signal
\begin{equation}
|\alpha_x,\vec{h}\rangle\in\{|\alpha,\vec{h}\rangle,|0,\vec{h}\rangle\in\mathcal{H}|\alpha\in\mathbb{R}\},
\end{equation}
one bit message is encoded depending on the existence of a coherent state. Whereas, in BPSK signal
\begin{equation}
|\alpha_x,\vec{h}\rangle\in\{|\alpha,\vec{h}\rangle,|-\alpha,\vec{h}\rangle\in\mathcal{H}|\alpha\in\mathbb{R}\},
\end{equation}
one bit message is encoded in a phase of a coherent state. Based on NS-CS, the Helstrom bounds of OOK and BPSK signals are given by
\begin{eqnarray}
P_{hel}(|\alpha,\vec{h}\rangle,|0,\vec{h}\rangle|q_0,q_1)=\frac{1+\sqrt{1-4q_0q_1\left\{h_0(\alpha^2)\right\}^2}}{2},\label{HB_OOK}
\end{eqnarray}
and
\begin{eqnarray}
P_{hel}(|\alpha,\vec{h}\rangle,|-\alpha,\vec{h}\rangle|q_0,q_1)=\frac{1+\sqrt{1-4q_0q_1\left[\sum_{n\in\mathbb{Z}^+\cup\{0\}}(-\alpha^2)^n\left\{h_n(\alpha^2)\right\}^2\right]^2}}{2},\label{HB_BPSK}
\end{eqnarray}
respectively. {We consider BPSK signal instead of OOK signal, since the Helstrom bound of the BPSK signal is larger than that of the OOK signal. We numerically show this in Fig.2 of Section 3.} 

In the next section, we propose the optimal measurement for discriminating BPSK signal composed of NS-CS. 

\subsection{Optimal Projective Measurement with Non-Standard Schr\"{o}dinger Cat States}

It is known that the optimal measurement for discriminating binary GS-CS can be described by a projective measurement composed of two orthogonal Schr\"{o}dinger cat state \cite{c.w.helstrom2}. Here, the optimal measurement is described by a projective measurement composed of two orthogonal Schr\"{o}dinger cat states
\begin{equation}
|\mathcal{C}_\pm\rangle:=\mathcal{N}_{\alpha_0,\alpha_1}^{(\pm)}(|\alpha_0\rangle\pm|\alpha_1\rangle),
\end{equation} 
where $\mathcal{N}_{\alpha_0,\alpha_1}^{(\pm)}$ is the normalization constant. Here, we define two orthogonal non-standard Schr\"{o}dinger cat states
\begin{equation}
|\mathcal{C}_{\pm,\vec{h}}\rangle:=\mathcal{N}_{\alpha_0,\alpha_1,\vec{h}}^{(\pm)}(|\alpha_0,\vec{h}\rangle\pm|\alpha_1,\vec{h}\rangle),\label{schrodinger_cat}
\end{equation}
where $\vec{h}$ is a tuple of real-valued functions defined in Definition 1. The following theorem show that the optimal measurement for discriminating NS-CS states in Eq.~(\ref{general_form_coherent_state}) is described by a projective measurement composed of the states in Eq. (\ref{schrodinger_cat}).
\\ \\
\textbf{Theorem 1.} Let us consider a projective measurement defined by
\begin{equation}
\Pi_y=|\pi_y\rangle\langle\pi_y|, \ \ y\in\{0,1\},\label{projective_measurement}
\end{equation}
where the orthonormal vectors $|\pi_0\rangle$ and $|\pi_1\rangle$ 
\begin{eqnarray}
|\pi_0\rangle:=\cos(\xi)|\mathcal{C}_+\rangle+e^{i\zeta}\sin(\xi)|\mathcal{C}_-\rangle,\label{basis9}
\end{eqnarray}
and
\begin{eqnarray}
|\pi_1\rangle:=\sin(\xi)|\mathcal{C}_+\rangle-e^{i\zeta}\cos(\xi)|\mathcal{C}_-\rangle,\label{basis}
\end{eqnarray}
respectively. For arbitrary $\alpha_0$ and $\alpha_1\in\mathbb{R}$ and prior probabilities $q_0$ and $q_1$, there exist $\xi$ and $\zeta$ that the maximal success probability $P_s(|\alpha_0\rangle,|\alpha_1\rangle|q_0,q_1)$ with respect to the projective measurement composed of $|\pi_0\rangle$ and $|\pi_1\rangle$ equals to the Helstrom bound.
\\ \\
\textit{Proof.} From Eqs. (\ref{projective_measurement}) and (\ref{basis}), maximal success probability is derived by
\begin{eqnarray}
&&P_{dir}(|\alpha_0,\vec{h}\rangle,|\alpha_1,\vec{h}\rangle|q_0,q_1)=q_0|\langle\pi_0|\alpha_0,\vec{h}\rangle|^2+q_1|\langle\pi_1|\alpha_1,\vec{h}\rangle|^2\nonumber\\
&&=\frac{1}{2}+\frac{(q_0-q_1)\langle\alpha_0,\vec{h}|\alpha_1,\vec{h}\rangle^2}{2}\cos(2\xi)+\frac{\sqrt{1-\langle\alpha_0,\vec{h}|\alpha_1,\vec{h}\rangle^2}}{2}\cos(\zeta)\sin(2\xi)\nonumber\\
&&\le \frac{1}{2}+\frac{(q_0-q_1)\langle\alpha_0,\vec{h}|\alpha_1,\vec{h}\rangle^2}{2}\cos(2\xi)+\frac{\sqrt{1-\langle\alpha_0,\vec{h}|\alpha_1,\vec{h}\rangle^2}}{2}\sin(2\xi).\label{succ_ineq}
\end{eqnarray}
Since $\alpha_0,\alpha_1\in\mathbb{R}$ are considered, $\langle\alpha_0,\vec{h}|\alpha_1,\vec{h}\rangle\in\mathbb{R}$ holds. Moreover, if $\zeta=0$, the inequality Eq. (\ref{succ_ineq}) becomes an equality. 

Suppose that $\xi$ takes the form:
\begin{equation}
\xi=\frac{1}{2}\arctan\left\{\frac{\sqrt{1-\langle\alpha_0,\vec{h}|\alpha_1,\vec{h}\rangle^2}}{(q_0-q_1)\langle\alpha_0,\vec{h}|\alpha_1,\vec{h}\rangle}\right\}.
\end{equation}
Then, Eq. (\ref{succ_ineq}) also takes the form equal to the Helstrom bound.\fbox

\\ \\
\noindent \textit{Remark 2.} The optimal projective measurement proposed in Theorem 1 requires technical capability for implementing the non-standard Schr\"{o}dinger cat states, whose experimental realization is still unknown. In the next section, we propose another near-optimal measurement, which can be possibly implemented in a near future.

\begin{figure}
\centering
\includegraphics[scale=0.7]{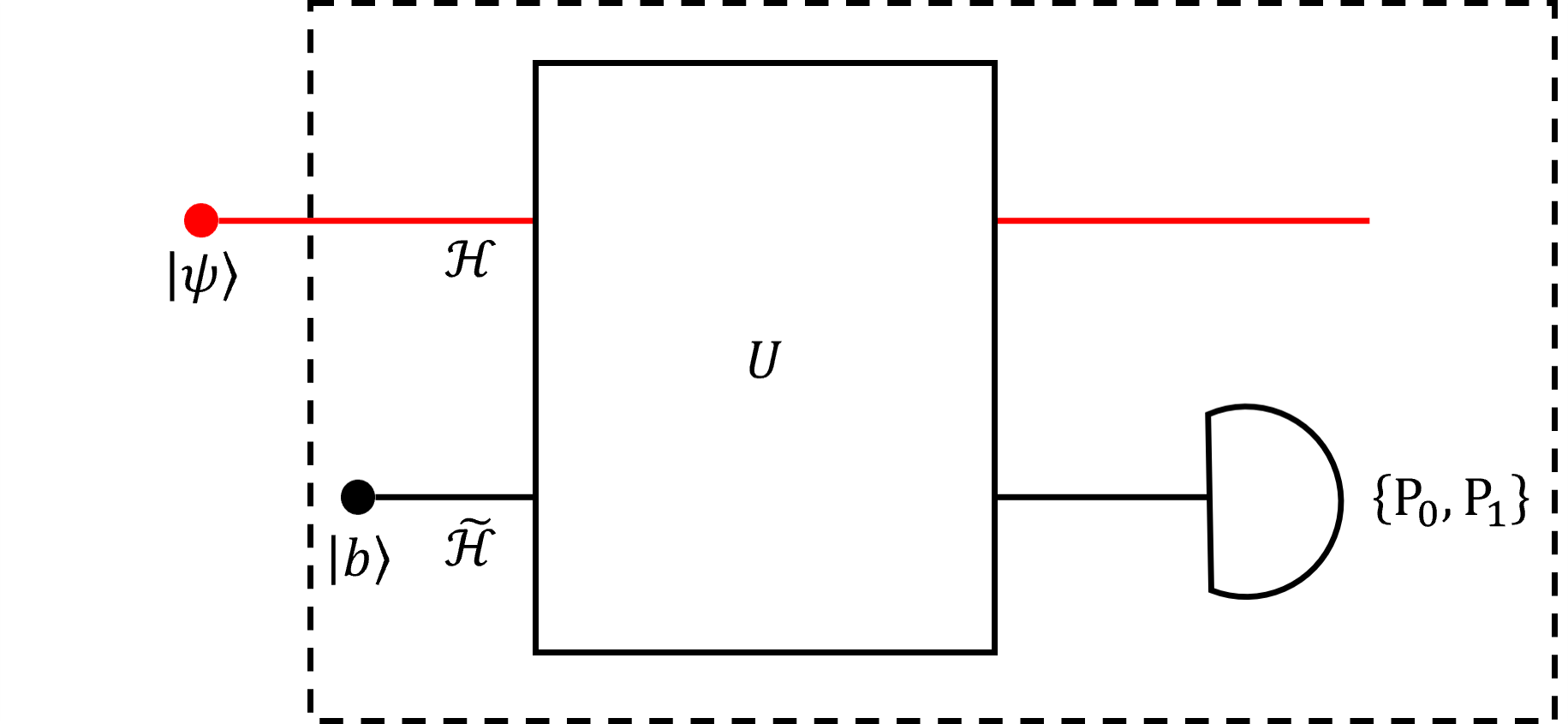}
\caption{Schematic of an indirect measurement. Here, $|b\rangle$ is the state of the auxiliary system $\widetilde{\mathcal{H}}$, $U$ is the unitary operator on $\mathcal{H}\otimes\widetilde{\mathcal{H}}$, and $\{\mathrm{P}_0,\mathrm{P}_1\}$ is a projective measurement on $\widetilde{\mathcal{H}}$.}
\end{figure}
\subsection{Optimal Indirect Measurement with Jaynes-Cummings Model}

\subsubsection{Success Probability of Indirect Measurement}
Recently, minimum error discrimination between binary pure states can be realized by an indirect measurement\cite{r.han}. This result is important because a rank-1 projective measurement is difficult to realize in some quantum systems including various continuous variable quantum lights. In other word, the indirect measurement needs to be considered for performing quantum state discrimination in such quantum systems. 

According to a mathematical framework\cite{m.ozawa}, an indirect measurement performed on a Hilbert space $\mathcal{H}$ is described by
\begin{equation}
\Xi:=\{\widetilde{\mathcal{H}},\sigma,\mathrm{P},U\},\label{statistical_realization}
\end{equation}
where $\widetilde{\mathcal{H}}$ is the auxiliary Hilbert space, $\sigma=|b\rangle\langle b|$ is the initial auxiliary state on $\widetilde{\mathcal{H}}$, $\mathrm{P}=\{\mathrm{P}_0,\mathrm{P}_1\}$ with
\begin{equation}
\mathrm{P}_y=|\pi_y\rangle\langle\pi_y|, \ \ |\pi_y\rangle\in\widetilde{\mathcal{H}},\label{projector}
\end{equation}
is the projective measurement performed on $\widetilde{\mathcal{H}}$. and $U$ is the unitary operator defined on a composite Hilbert space $\mathcal{H}\otimes\widetilde{\mathcal{H}}$. Fig.1 illustrates the indirect measurement consist of $\sigma$, $\mathrm{P}$ on $\widetilde{\mathcal{H}}$ and $U$ on $\mathcal{H}\otimes\widetilde{\mathcal{H}}$.

For a pure state $|\psi\rangle\in\mathcal{H}$, $U$ satisfies the following relation\cite{e.r.loubenets,e.r.loubenets2}
\begin{equation}
U(|\psi\rangle\otimes|b\rangle)=\sum_{y}K_y|\psi\rangle\otimes|\pi_y\rangle,\label{unitary_representation}
\end{equation}
where $K_y$ is a Kraus operator. From Eq. (\ref{unitary_representation}), the physical interpretation of $K_y$ is obtained by
\begin{equation}
K_y=\langle\pi_y|U|b\rangle_{\widetilde{\mathcal{H}}}.\label{Kraus_operator}
\end{equation}
By using Eq. (\ref{Kraus_operator}), success probability for discriminating NS-CS via the indirect measurement can be rewritten as
\begin{equation}
P_s(|\alpha_0,\vec{h}\rangle,|\alpha_1,\vec{h}\rangle|q_0,q_1)=q_0\langle \alpha_0,\vec{h}|K_0^\dagger K_0|\alpha_0,\vec{h}\rangle+q_1\langle \alpha_1,\vec{h}|K_1^\dagger K_1|\alpha_1,\vec{h}\rangle.\label{success_probability_indirect}
\end{equation} 

\subsubsection{Indirect Measurement with Jaynes-Cummings Model}
An indirect measurement considered in the present article uses a two-level atom as an auxiliary quantum system $\widetilde{\mathcal{H}}$. We denote $\widetilde{\mathcal{H}}=\mathrm{span}\{|g\rangle,|e\rangle\}$ where $|g\rangle$ is the ground state and $|e\rangle$ is the excited state of the two-level atom. The initial state of the two-level atom in $\widetilde{\mathcal{H}}$ is defined by
\begin{equation}
\sigma:=|g\rangle\langle g|.\label{initial_auxiliary_state}
\end{equation} 

Each projector in Eq. (\ref{projector}) is constructed by orthonormal vectors:
\begin{eqnarray}
|\pi_0\rangle:=\cos(\theta)|g\rangle+e^{i\phi}\sin(\theta)|e\rangle,\label{orthonormal_vector0}
\end{eqnarray}
and
\begin{eqnarray}
|\pi_1\rangle:=\sin(\theta)|g\rangle-e^{i\phi}\cos(\theta)|e\rangle.\label{orthonormal_vector}
\end{eqnarray}

In this article, we design the unitary operator $U$ in Eq. (\ref{statistical_realization}) using Jaynes-Cummings model\cite{e.t.jaynes}. In Jaynes-Cummings model, an interaction between a light and a two-level atom is described by a Hamiltonian on the composite Hilbert space $\mathcal{H}\otimes\widetilde{\mathcal{H}}$
\begin{equation}
H(t):=H_0+H_{int}(t),\label{entire_hamiltonian}
\end{equation}
where $H_0$ and $H_{int}$ are free Hamiltonian and interaction Hamiltonian defined by
\begin{eqnarray}
H_0:=\hbar\omega_L(a^\dagger a\otimes\mathbb{I}_{\widetilde{\mathcal{H}}})+\frac{1}{2}\hbar\omega_0(\mathbb{I}_{\mathcal{H}}\otimes\sigma_z),
\end{eqnarray}
and
\begin{eqnarray}
H_{int}(t):=\hbar\Omega(t)(a\otimes\sigma_++a^\dagger\otimes\sigma_-),\label{each_hamiltonian}
\end{eqnarray}
respectively. Here, $\omega_L$ is a light frequency, $\omega_0$ is a transition frequency of the two-level atom, and $\Omega(t)$ is a time-dependent interaction parameter. $a$ is an annihilation operator and $a^\dagger$ is a creation operator such that
\begin{equation}
a|n\rangle=\sqrt{n}|n-1\rangle, \ \ a^\dagger|n\rangle=\sqrt{n+1}|n+1\rangle.
\end{equation}
The Pauli operators are based on the ground and excited states of $\widetilde{\mathcal{H}}$ as 
\begin{equation}
\sigma_z:=|e\rangle\langle e|-|g\rangle\langle g|, \ \ \sigma_+:=|e\rangle\langle g|, \ \ \sigma_-=|g\rangle\langle e|.\label{Pauli}
\end{equation}

While the light and the two-level atom are resonating ($\omega_L=\omega_0$), Eq. (\ref{entire_hamiltonian}) is simplified by the interaction picture on the basis of $H_{0}$: 
\begin{equation}
\widetilde{H}_{int}(t)=\hbar\Omega(t)(a\otimes\sigma_++a^\dagger\otimes\sigma_-).
\end{equation}
To evaluate $U$ in Eq. (\ref{statistical_realization}), consider a time-dependent unitary operator $\widetilde{U}(t)$ as a solution of the time-dependent Schr\"{o}dinger equation:
\begin{equation}
i\hbar\frac{d\widetilde{U}(t)}{dt}=\widetilde{H}_{int}(t)\widetilde{U}.
\end{equation}
The solution of the time-dependent Schr\"{o}dinger equation is analytically given by \cite{m.namkung2}
\begin{equation}
\widetilde{U}(t):=\exp\left\{-i\widetilde{\Phi}(t)(a\otimes\sigma_++a^\dagger\otimes\sigma_-)\right\}.\label{total_U}
\end{equation}
Here, $\widetilde{\Phi}(t)$ is a time-dependent parameter:
\begin{equation}
\widetilde{\Phi}(t):=\int_0^t\Omega(\tau)d\tau.
\end{equation}
Finally, given that the interaction time is $T$, the unitary operator $U$ in Eq. (\ref{statistical_realization}) is defined by
\begin{equation}
U:=\widetilde{U}(T).\label{unitary_construction}
\end{equation}

From Eqs. (\ref{initial_auxiliary_state}), (\ref{orthonormal_vector0}), (\ref{orthonormal_vector}) and (\ref{unitary_construction}) together with Eq. (\ref{Kraus_operator}), the Kraus operators of the indirect measurement in Eq. (\ref{Kraus_operator}) can be rewritten as \cite{m.namkung2}
\begin{eqnarray}
K_0:=\cos(\theta)\cos\{\Phi|a|\}-ie^{-i\phi}\sin(\theta)\sum_{k\in\in\mathbb{Z}^+\cup\{0\}}\frac{(-1)^k}{(2k+1)!}\Phi^{2k+1}a|a|^{2k},\label{JC_Kraus_0}
\end{eqnarray}
and
\begin{eqnarray}
K_1:=\sin(\theta)\cos\{\Phi|a|\}+ie^{-i\phi}\cos(\theta)\sum_{k\in\in\mathbb{Z}^+\cup\{0\}}\frac{(-1)^k}{(2k+1)!}\Phi^{2k+1}a|a|^{2k},\label{Jaynes_Cummings_Kraus}
\end{eqnarray}
where $\Phi$ is a parameter defined by
\begin{equation}
\Phi:=\widetilde{\Phi}(T),
\end{equation} 
and $|a|:=\sqrt{a^\dagger a}$. 

From Eqs. (\ref{JC_Kraus_0}) and (\ref{Jaynes_Cummings_Kraus}) together with Eq. (\ref{success_probability_indirect}), the success probability that the indirect measurement discriminates the binary pure states is rewritten by
\begin{equation}
P_s(|\alpha_0\rangle,|\alpha_1\rangle|q_0,q_1)=q_0\sum_{n\in\mathbb{Z}^+\cup\{0\}}|\mathcal{F}_0(n)|^2+q_1\sum_{n\in\mathbb{Z}^+\cup\{0\}}|\mathcal{F}_1(n)|^2,\label{objective_function}
\end{equation}
where $\mathcal{F}_y:\mathbb{Z}^+\cup\{0\}\rightarrow\mathbb{R}^+$ is a real-valued function defined by
\begin{eqnarray}
\mathcal{F}_0(n):=\left|\alpha_0^nh_n(|\alpha_0|^2)\cos(\theta)\cos(\Phi\sqrt{n})-i\alpha_0^{n+1}h_{n+1}(|\alpha_0|^2)e^{-i\phi}\sin(\theta)\sin(\Phi\sqrt{n+1})\right|^2,\label{F0}
\end{eqnarray}
and
\begin{eqnarray}
\mathcal{F}_1(n):=\left|\alpha_1^nh_n(|\alpha_1|^2)\sin(\theta)\cos(\Phi\sqrt{n})+i\alpha_1^{n+1}h_{n+1}(|\alpha_1|^2)e^{-i\phi}\cos(\theta)\sin(\Phi\sqrt{n+1})\right|^2.\label{F1}
\end{eqnarray}
By using Eq. (\ref{F0}) and Eq. (\ref{F1}), the optimal success probability of the indirect measurement is described by
\begin{eqnarray}
P_{ind}(|\alpha_0,\vec{h}\rangle,|\alpha_1,\vec{h}\rangle|q_0,q_1):=\max_{(\theta,\phi,{\Phi})\in\mathbb{R}^3}\left\{q_0\sum_{n\in\mathbb{Z}^+\cup\{0\}}|\mathcal{F}_0(n)|^2+q_1\sum_{n\in\mathbb{Z}^+\cup\{0\}}|\mathcal{F}_1(n)|^2\right\}.\label{optimal_success_probability}
\end{eqnarray}

Since Eq. (\ref{objective_function}) is nonlinear, evaluating Eq. (\ref{optimal_success_probability}) becomes a nonlinear optimization problem. This implies that analytically finding the optimal parameters $(\theta,\phi,\Phi)$ is difficult. Fortunately, the optimization problem for evaluating Eq. (\ref{optimal_success_probability}) is not constrained. Therefore, we can consider the Powell method or the steepest ascent method to numerically obtain the maximal success probability.

\begin{figure}
\centering
\includegraphics[scale=0.45]{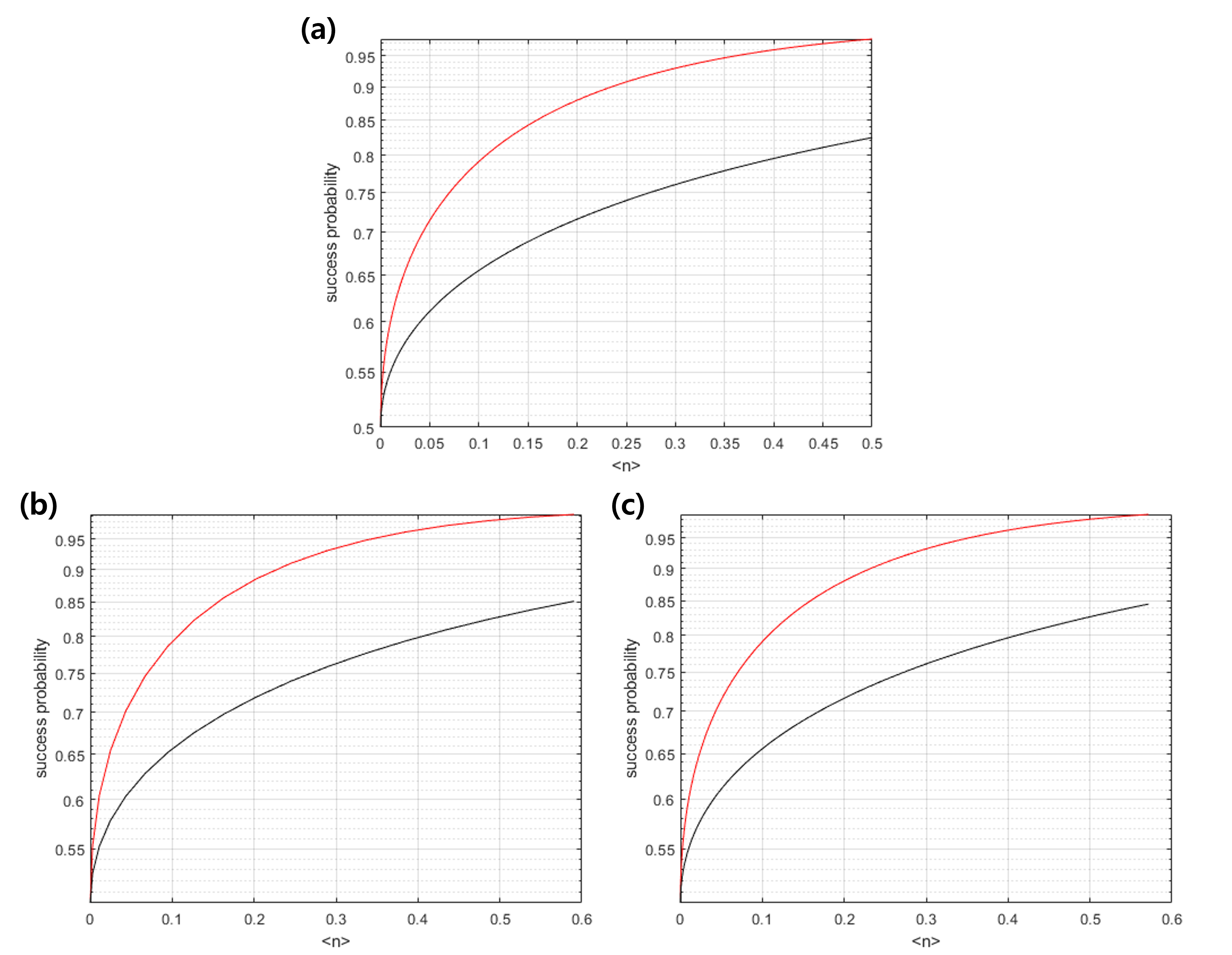}
\caption{{The Helstrom bounds between binary signals. Here, solid red lines show BPSK signal and solid black lines show OOK signal. (a) shows OS-CS with $n_j=3$, (b) shows BG-CS with $\chi=1/2$, and (c) shows mSG-CS, respectively.}}
\end{figure}

\subsubsection{Numerical Investigation of Optimal Success Probability}
Here, we numerically investigate the optimal success probability for discriminating NS-CS for the case of an optical spin coherent state (OS-CS), a Barut-Girardello coherent state (BG-CS)\cite{a.o.barut} and a modified Susskind-Glogower coherent state (mSG-CS)\cite{j.-p.gazeau}.
\begin{itemize}
\item In OS-CS, $h_n(u)$ in Eq. (\ref{general_form_coherent_state}) is expressed by\cite{a.m.perelomov}
\begin{eqnarray}
h_n(u)=\lambda_n(1+u)^{-\frac{n_j}{2}}, \ \ \lambda_n=\sqrt{\frac{n_j!}{n!(n_j-n)!}},
\end{eqnarray}
where $n_j$ is a positive integer. For $n>n_j$, every $h_n(u)$ equals to zero. 
\item In BG-CS, $h_n(u)$ in Eq. (\ref{general_form_coherent_state}) is expressed by\cite{a.o.barut}
\begin{eqnarray}
h_n(u)=\frac{\lambda_n}{\sqrt{\mathcal{N}_{BG}(u)}}, \ \ \lambda_n=\sqrt{\frac{\Gamma(2\chi)}{n!\Gamma(2\chi+n)}}.
\end{eqnarray}
Here, $\chi$ is a real number such that $\chi\ge1/2$, $\Gamma(\cdot)$ is a Gamma function, and $\mathcal{N}_{BG}(u)$ is a normalization function defined by
\begin{equation}
\mathcal{N}_{BG}(u):=\Gamma(2\chi)u^{\frac{1}{2}-\chi}I_{2\chi-1}(2\sqrt{u}),
\end{equation}
where $I_{\nu}(\cdot)$ is a modified Bessel function.
\item In mSG-CS, $h_n(u)$ in Eq. (\ref{general_form_coherent_state}) is expressed by\cite{j.-p.gazeau}
\begin{equation}
h_n(u)=\sqrt{\frac{n+1}{\mathcal{N}_{mSG}(u)}}\frac{1}{u^{\frac{n+1}{2}}}J_{n+1}(2\sqrt{u}).
\end{equation}
Here, $J_n(\cdot)$ is a first kind of Bessel function, and $\mathcal{N}_{mSG}(u)$ is a normalization function defined by
\begin{eqnarray}
\mathcal{N}_{mSG}(u):=\frac{1}{u}\left[2u\{J_0(2\sqrt{u})\}^2-\sqrt{u}J_0(2\sqrt{u})J_1(2\sqrt{u})+2u\{J_1(2\sqrt{u})\}^2\right].
\end{eqnarray}
\end{itemize}
\begin{figure}
\centering
\includegraphics[scale=0.3]{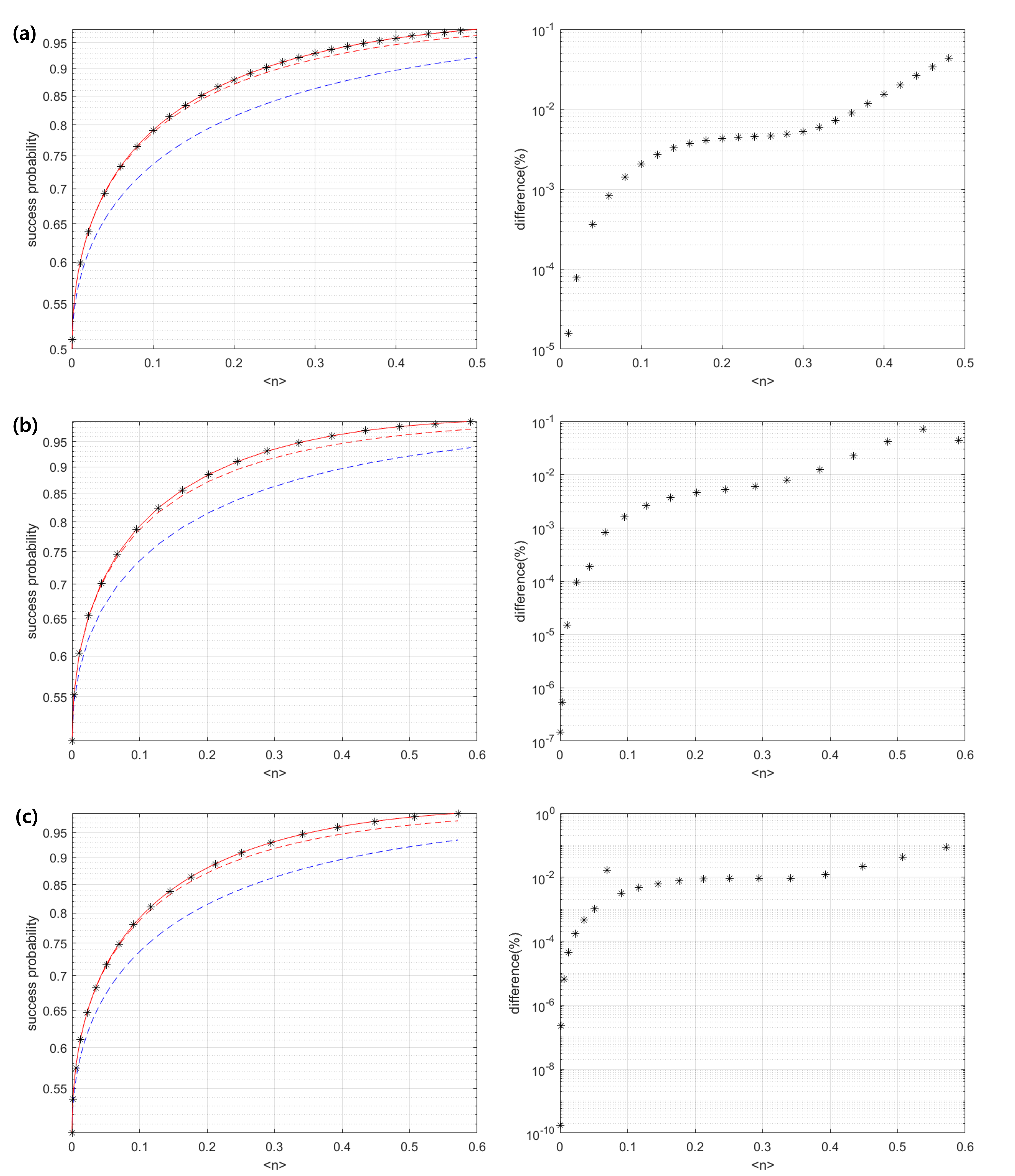}
\caption{(Left) Optimal success probabilities for discriminating BPSK composed of one of three NS-CS, (Right) percentages of difference between an optimal success probability and the Helstrom bound. (a) shows OS-CS with $n_j=3$, (b) shows BG-CS with $\chi=1/2$, and (c) shows mSG-CS. Also, in the left graphs, solid red line depicts the Helstrom bound between binary NS-CS, dashed red line depicts the Helstrom bound between binary S-CS, and dashed blue line depicts the shot noise limit of S-CS\cite{i.a.burenkov0,k.tsujino}.}
\end{figure}
{In Fig.2, the Helstrom bounds between binary signals are illustrated. Here, solid red lines show BPSK signal and solid black lines show OOK signal. Fig.2(a) shows OS-CS with $n_j=3$, Fig.2(b) shows BG-CS with $\chi=1/2$, and Fig.2(c) shows mSG-CS, respectively. According to Fig.2, the values of solid red lines are larger than those of the solid black lines, which means the quantum communication can be enhanced by the BPSK signals rather than OOK signals.}

{Therefore, we focus on the optimal implementation of the quantum communication with BPSK signal.} In Fig.3, the graphs in the left column depict optimal success probabilities for discriminating BPSK composed of one of three NS-CS. The graphs in the right column depict difference between an optimal success probability and the Helstrom bound defined by
\begin{eqnarray}
&&\delta(|\alpha_0,\vec{h}\rangle,|\alpha_1,\vec{h}\rangle|q_0,q_1)\\
&&:=100\times\frac{P_{hel}(|\alpha_0,\vec{h}\rangle,|\alpha_1,\vec{h}\rangle|q_0,q_1)-P_{ind}(|\alpha_0,\vec{h}\rangle,|\alpha_1,\vec{h}\rangle|q_0,q_1)}{P_{hel}(|\alpha_0,\vec{h}\rangle,|\alpha_1,\vec{h}\rangle|q_0,q_1)},\nonumber
\end{eqnarray}
where the factor 100 is to normalize in percentage. Fig.3(a) shows OS-CS with $n_j=3$, Fig.3(b) shows BG-CS with $\chi=1/2$, and Fig.3(c) shows mSG-CS. For the graphs in the left column of Fig.3, solid red line depicts the Helstrom bound between binary NS-CS, dashed red line depicts the Helstrom bound between binary S-CS, and dashed blue line depicts the shot noise limit of S-CS\cite{i.a.burenkov0,k.tsujino}. Black points are numerical values of Eq. (\ref{optimal_success_probability}). 

\begin{itemize}
\item According to Fig.3, the optimal success probability of the indirect measurement in Eq. (\ref{optimal_success_probability}) nearly achieves the Helstrom bound if the mean photon number satisfies $0<\langle n\rangle<0.6$. Fig.3 also shows that difference between Eq. (\ref{optimal_success_probability}) and the Helstrom bound is less than $10^{-5}$\% if $\langle n\rangle$ is small enough. 
\item If $\langle n \rangle$ is large, the difference between Eq. (\ref{optimal_success_probability}) and the Helstrom bound increases. Nevertheless, the difference is less than $0.1$\% in region of $0<\langle n\rangle<0.6$. 
\end{itemize}
This implies that the indirect measurement based on Jaynes-Cummings model can perform the optimal quantum communication enhanced by sub-Poissonianity of NS-CS. Because Jaynes-Cummings model describes the interaction between the light and the two-level atom, our scheme has potential to be implemented in the near future.

\section{Conclusions}
In the present article, we proposed optimal strategies for quantum communication enhanced by non-standard coherent states. 
\begin{itemize}
\item We have analytically shown that optimal measurement for the enhanced quantum communication is described by a projective measurement composed of orthogonal non-standard Schr\"{o}dinger cat states. 
\item We have numerically shown that an indirect measurement based on the Jaynes-Cummings model performs the near-optimal quantum-enhanced communication. Since the Jaynes-Cummings model describes the interaction between the light and the two-level atom, it can be realized in the future.
\end{itemize} 

The success probability for discriminating the BPSK signal can be larger than that of the OOK signal and this justifies the reason of concerning BPSK signal instead of OOK signal. As Jaynes-Cummings model is known to be suitable in the field of quantum computation \cite{j.i.cirac}, our result provides a useful reference in the field of quantum computation and communication.

\section*{Acknowledgements}
This work was supported by Quantum Computing Technology Development Program (NRF2020M3E4A1080088) through the National Research
Foundation of Korea (NRF) grant funded by the Korea government (Ministry of Science and ICT). MN and JSK appreciate Dr. Donghoon Ha and Prof. Kwang Jo Lee for their insightful discussions.

\end{document}